%% file: ISQEDsamp.tex
\documentclass[10pt,twocolumn]{IEEEtran} 

\usepackage[pdftex]{graphicx}
\usepackage{xcolor,colortbl}
\usepackage{cite}
\usepackage{amsmath}
\usepackage[flushleft]{threeparttable}
\usepackage{textcomp}
\usepackage[none]{hyphenat}
\usepackage{enumitem}
\usepackage{rotating}
\usepackage{multirow}
\usepackage{graphicx}
\usepackage{authblk}

\usepackage{fancyhdr}

\begin{document}

\title{Minimizing Classification Energy of Binarized Neural Network Inference for Wearable Devices} 


\author{\large Morteza Hosseini, Hirenkumar Paneliya, Uttej Kallakuri, Mohit Khatwani, and Tinoosh Mohsenin}
\affil{\normalsize Department of Computer Science and Electrical Engineering, University of Maryland, Baltimore County}

\maketitle
\thispagestyle{empty}\pagestyle{empty}

\begin{abstract}
In this paper, we propose a low-power hardware for efficient deployment of binarized neural networks (BNNs) that have been trained for physiological datasets. BNNs constrain weights and feature-map to $1$ bit, can pack in as many 1-bit weights as the width of a memory entry provides, and can execute multiple multiply-accumulate (MAC) operations with one fused bit-wise xnor and population-count instruction over aligned packed entries. Our proposed hardware is scalable with the number of processing engines (PEs) and the memory width, both of which adjustable for the most energy efficient configuration given an application. We implement two real case studies including Physical Activity Monitoring and Stress Detection on our platform, and for each case study on the target platform, we seek the optimal PE and memory configurations. Our implementation results indicate that a configuration with a good choice of memory width and number of PEs can be optimized up to 4$\times$ and 2.5$\times$ in energy consumption respectively on Artix-7 FPGA and on 65nm CMOS ASIC implementation. We also show that, generally, wider memories make more efficient BNN processing hardware. To further reduce the energy, we introduce Pool-Skipping technique that can skip at least 25$\%$ of the operations that are accompanied by a Max-Pool layer in BNNs, leading to a total of 22$\%$ operation reduction in the Stress Detection case study. Compared to the related works using the same case studies on the same target platform and with the same classification accuracy, our hardware is respectively 4.5$\times$ and 250$\times$ more energy efficient for the Stress Detection on FPGA and Physical Activity Monitoring on ASIC, respectively.



\end{abstract}


\pagestyle{fancy}
\rhead{\textcolor{black}{\textit{\textbf{\small ISQED'19, March 2019, Santa Clara, CA}}}}

\section{\bf{Introduction}}

In recent years, the demand for using wearable technology has increased dramatically with the advancement of technology and the growth of the Internet of Things (IoT). Wearable devices are extending their applications in many diverse domains from consumer electronics, such as smartwatches and activity trackers for monitoring one's fitness and wellness, to medical devices that track a patient's vital physiological signs such as heart rate and brain activity. Such devices usually process real-time data, read from multimodal sensors continuously, and suffer from resource-bound and limited battery budget due to their small size, online monitoring, and portability. Therefore, minimizing the power dissipation of these devices while meeting real-time requirements is a subject of interest \cite{kim2017kernel, jafari2018sensornet, page2015utilizing, abtahi2017accelerating, abtahi2018accelerating}.

Deep Neural Networks (DNNs) are among the effective machine learning (ML) approaches that are employed majorly in the processing units of such devices. DNNs can extract featured information from raw multimodal time-series signals without much prior knowledge of the signal \cite{jafari2018sensornet}. However, DNNs with high-precision weights consume high-energy operations and require large blocks of memories for storing and processing the DNN model. To tackle the problem of memory and power consumption, weight pruning and weight quantization techniques \cite{chen2015compressing, shen2018least, han2015deep} as well as low precision weight networks \cite{courbariaux2016binarized,courbariaux2015binaryconnect,rastegari2016xnor} have been proposed in recent years as effective approaches to compressing DNN models. \cite{courbariaux2016binarized}, \cite{courbariaux2015binaryconnect}, and \cite{rastegari2016xnor} proposed respectively BinaryConnect (BC), Binarized Neural Network (BNN), and Binary-Weight-Networks (BWN), all of which constrain every weight of the neural network to either $-1$ or $+1$, that can be stored in the smallest possible unit of memory, i.e. 1 bit ($0$ for $-1$ and $1$ for $+1$), thereby allowing to pack as many weights in as a memory entry can accommodate. In BNNs, not only the weights but the feature-map is also constrained to $-1$ or $+1$, thus resulting in bit-wise operations between each neuron output and every synapse weight. Due to the minimal aspects of binary weight DNNs, they have gained significant attention recently. BNNs work well on small datasets \cite{courbariaux2016binarized}, and will be shown in this work that they can be employed on physiological time-series data as well. 

Time-series data classification in ML applications is conventionally processed with techniques such as Dynamic Time Warping (DWT), Recurrent Neural Networks (RNN), and Long Short-Term Memory (LSTM) \cite{karim2018lstm}. These techniques have complicated algorithms and hardware implementation, and mainly deal with complex input data such as text or speech. DNNs, nevertheless, have also been used in the classification of time-series data \cite{jafari2018sensornet,zheng2014time}. In this approach, the real-time data is cut and buffered into frames of samples that may or may not contain specific patterns, e.g. seizure pattern, that annotate the frame with specific labels. The frame of temporal data can be multi-channel and can be used as raw data or processed in the frequency domain to feed a DNN, or represented as an image-like frame whose sequence of image rows follow the sequence of sensors that pick data, e.g. EEG sensors on the skull can represent the brain activity spatially and temporally \cite{soleymani2016analysis}.

In this paper, a low-power scalable hardware is proposed for multimodal physiological data classification that uses BNNs. In its Verilog HDL, the number of PEs and memory width can be adjusted to fit the minimal energy consumption given an application. The hardware is designed to be minimal, and use resource sharing and on-chip memories at its finest. 

The main contributions of this paper include:
\begin{itemize}
\item Train two physiological case studies including Physical Activity Monitoring, and Stress Detection for BNNs
\item Improve BNN accuracy by increasing the number of parameters to achieve the same accuracy as full-precision
\item Propose a scalable parallel hardware for BNN inference that can be configured for different number of PEs, and memory width, both of which adjustable for minimal classification energy given a case study.
\item Propose an operation skipping technique, referred to as \textit{Pool-Skipping}, for Max-Pool layer in BNNs that can skip more than 25\% of effectless operations accompanied with MP and 22\% of the BNN operations in total.

\item Analyze hardware parameters, including energy vs memory width, and energy vs number of PEs and find the optimal configuration per case study on FPGA and on ASIC.
\item Synthesis and Implementation of the platform on FPGA and post place-and-route ASIC layout in 65nm CMOS technology, and provide the energy and time reports of the two case studies.
\end{itemize}



\section{\bf{Related Work}}
Several multimodal data classification methods have been proposed recently. \cite{jafari2018sensornet} proposed a complexity-reduced Deep Convolutional Neural Network (DCNN) for physiolog- ical case studies and a 16-bit hardware platform for their inference. \cite{zheng2014time} proposed an architecture that uses a CNN per variable and tested their CNN on Congestive Heart Failure Detection and on Physical Activity Monitoring datasets with detection accuracies of respectively 94\% and 93\%. \cite{li2017concurrent} proposed a scalable system that addresses concurrent activity recognition. Their model extracts temporal and spatial features from a multimodal sensory data by means of a 7 layer CNN followed by an LSTM and is tested on three activity datasets. \cite{yao2017deepsense} introduced DeepSense which is a deep learning framework that integrates DNNs and RNNs for addressing feature customization challenges in sensory datasets. \cite{radu2016towards} investigated the opportunity to use deep learning for identifying nonintuitive features from cross-sensor correlations by means of an RBM. In \cite{kim2017kernel} a kernel decomposing scheme in binary-weight networks is proposed that skips redundant computations and achieves 22\% energy reduction on image classification. BiNMAC is proposed in \cite{GLSVLSI-ali} which is a programmable manycore accelerator for BNNs designed for physiological and Image processing case studies.


\section{\bf{Binarized Neural Networks}}
BNNs constrain both weights and the feature-map to either $-1$ or $+1$. Similar to full-precision DNNs, BNNs consist of layers equivalent to those of the standard DNNs, albeit implemented rather differently; for the first layer of BNN, data from each input channel are in full-precision values, but model weights of the BNN are trained and constrained to either $-1$ or $+1$. Therefore, Multiply-Accumulation (MAC) operations are replaced by a series of ADDs/SUBs. For the next layers, all input features are binarized using binary activation functions (AF) and packed inside registers. Therefore, a MAC operation between two vectors that hold such packed weights can be performed by summing (considering $0$ as $-1$) over the result of the bit-wise $xnor$ of the two vectors. The summation over the bits of a register is referred to as $population-count$ and denoted by $pcnt$ in this work.


In order to apply the $AF$ to the real-valued variables and transform them to either $-1$ or $+1$, deterministic binarization function is proposed for AF by \cite{courbariaux2015binaryconnect} which is essentially a $sign$ function as shown in equation (\ref{eq:eq-deter}), where $x$ is the real-valued variable and $x_b$ is its binarized value.

    \begin{equation} 
      x_{b} \text{ = sign($x$) = }\begin{cases}
                   \,+1\qquad if x\geq 0\\
                  \,-1\quad otherwise,
                \end{cases}
    \label{eq:eq-deter}
    \end{equation}
Thus, the feed-forward path of a layer, that is traditionally denoted with $Y=AF(W.X+B)$ in standard DNNs, can be formulated for BNNs by equation (\ref{eq:BNNoperation1}): 

\begin{equation} 
Y = sign( pcnt( xnor (W,X)) )
\label{eq:BNNoperation1}
\end{equation}
where \textit{X} and \textit{W} are respectively tensors of input (feature-map) and weights with bias ($B$) implicitly included, $xnor$ performs bit-wise $xnor$ operations on rows from \textit{W} and columns from \textit{X}, $pcnt$ sums over the result of this bit-wise operation, and \textit{sign} is the AF applied to the generated scalar value from $pcnt$. As explained earlier, from a hardware point of view it is more efficient for both binarized weights and feature-map to be packed in memory entries. For example, one row of a binarized $W_{128\times128}$ can be stored in 4 memory entries of a 32-bit machine, and the whole matrix can be packed inside 512 entries (2KB). Therefore, practically in an M-bit computing machine, the population-count part in equation (\ref{eq:BNNoperation1}) should be performed as per every chunk of aligned bit-wise $xnor$. Hence, before applying the activation function the population-count needs to be carried out several times until the completion of fetching all patch elements that contribute to generating one output value. This accumulation, before applying the $sign$ AF, can be denoted by the following equation \cite{courbariaux2016binarized}:

\begin{equation} 
temp += pcnt( xnor (W_{M},X_{M}))
\label{eq:BNNoperation2}
\end{equation}
where $W_{M}$ and $X_{M}$ are chunks of $W$ and $X$ that are sequentially read from packed memory entries of an M-bit machine until they are fetched completely. Only then one can infer $Y=sign(temp)$ from equations (\ref{eq:BNNoperation1}) and (\ref{eq:BNNoperation2}).


For the last layer, weights can either be constrained to $-1$ or $+1$, that would be treated similarly as previous layers, or trained with full-precision values that, with respect to the binarized activation values coming from the previous layer would convert the MAC operation into ADD/SUB over the weight values, analogous to the first layer. Binarizing the last layer of BNNs can sometimes result in failure and therefore different techniques have been proposed to deal with the last layer \cite{tang2017train
, umuroglu2017finn}. We use high-precision (16-bit) weights for the last layer of the first case study in this paper because it resulted in a stabilized BNN accuracy during trainaing over 100 epochs.


 

Max-Pool (MP) is a frequently-employed layer in DNNs that sub-samples the feature-map, and is traditionally placed before $AF$. It can be shown that if $AF$ is a non-descending function of its input, then placing $MP$ after $AF$ results in equivalent outcome. In BNNs therefore:

\begin{equation} 
MP(sign(Y))=sign(MP(Y))
\label{eq:BNNoperationMP}
\end{equation}

Placing $MP$ after $AF$ layer in BNNs has two advantages: 1) a bulky comparator that would have to operate on the real-valued results from the previous Conv/FC layer in order to execute $MP$ before $AF$, can be converted to an $or$ gate if $MP$ follows $AF$ with binarized output. Because, 
if $y_i$ $\in$ $\{$$0$, $1$$\}$, then $max$($y_1$, $y_2$, ..., $y_n$) = $or$($y_1$, $y_2$, ...,$y_n$), where $y_i$ and $n$ are elements and the pool size of the MP 2) as soon as one of the elements of the pool, at which the $MP$ seeks the maximum, is equal to the maximum possible value ($1$ in BNNs), the pool exploring can be asserted as `done', and the rest of the exploration in the pool can be skipped, that is conducive to skipping of large portion of computation imposed from the preceding Conv/FC layer. We refer to this property as \textit{Pool-Skipping} and in Section VI show that it can skip up to 22\% of the total operations for the first case study that has 3 $MP$ layers.

Since in Neural Networks one operation is usually considered as either one multiplication or one addition, therefore each time either a bit-wise $xnor$ or a $pcnt$ is executed, the number of carried out operations is equal to the number of vector bits to be executed. Consequently in a BNN process, by fusing and well pipelining the $pcnt$ and bit-wise $xnor$ operations in an M-bit machine, one can achieve 2$\times$M operations per execution of equation \ref{eq:BNNoperation2}. This implies that the performance of an M-bit BNN processing machine can be proportional to M, which decides its registers, data bus, memory width, etc. We will evaluate this feature in Section VI.

\section{\bf{BNN for Physiological Case Studies}}

BNN models can compress the model size by 8 to 64 times, at the expense of some accuracy loss. The accuracy loss can usually be compensated by augmenting the BNN model size by 2-11$\times$ \cite{umuroglu2017finn}. 
In order to present a fair comparison between our hardware and related works, in this section, we introduce two case studies that were used in \cite{jafari2018sensornet} and \cite{GLSVLSI-ali}. We use Torch framework from \cite{courbariaux2016binarized} to train the BNN models and to grab the weights for inference on hardware.

\subsection{Case study 1: Stress Detection}

\begin{figure}
\begin{center}
\includegraphics[width=3.5in]{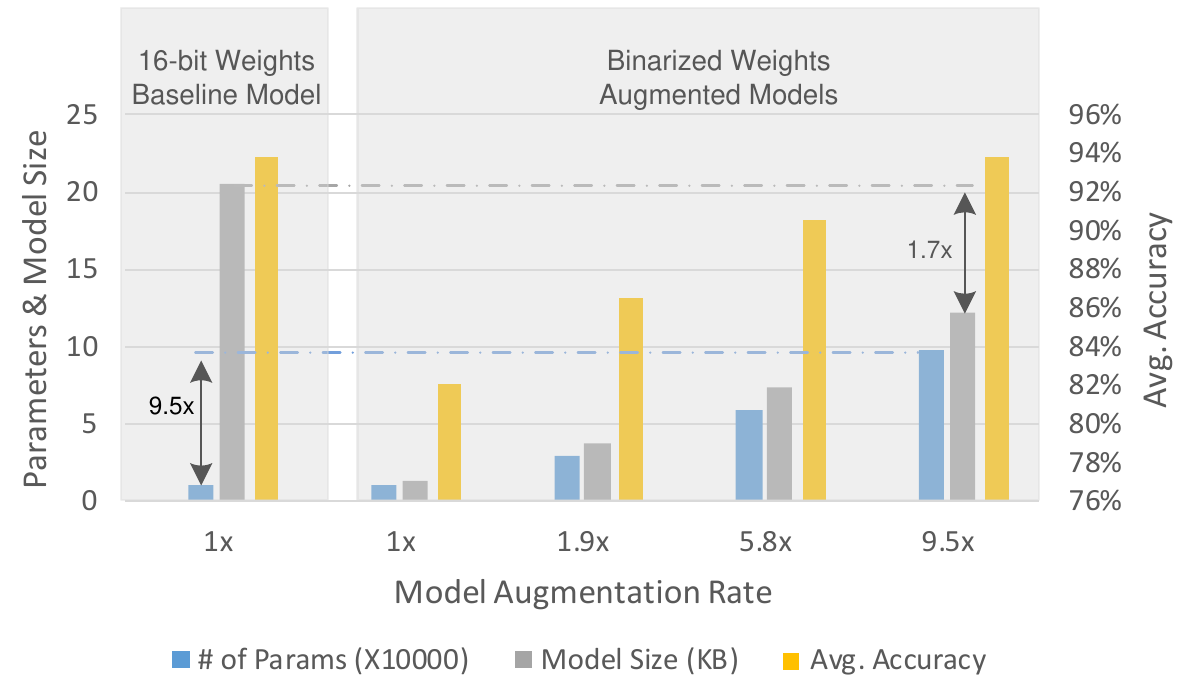}
\end{center}
\caption{Augmentation impact on the number of parameters, model size, and accuracy from a 16-bit baseline model for Stress Detection to BNN models 
\label{fig:Augmentation}}
\end{figure}

\begin{figure}
\begin{center}
\includegraphics[width=3.3in]{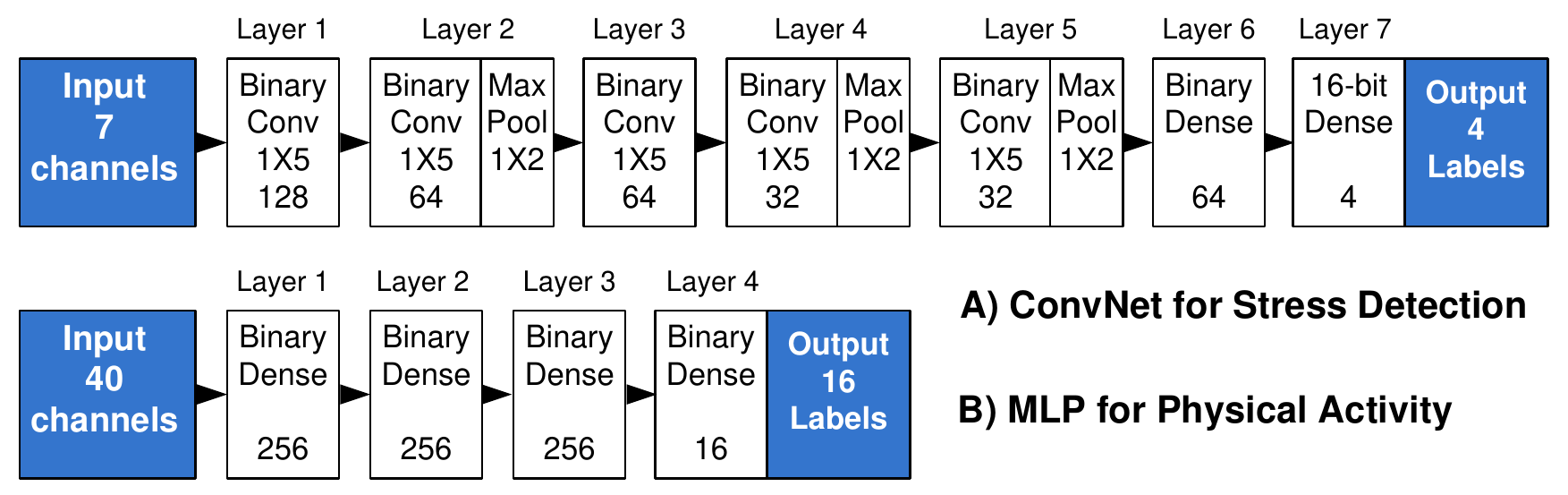}
\end{center}
\caption{BNN architectures of the two case studies used in this paper
\label{fig:case_models}}
\end{figure}

The Stress Detection dataset \cite{birjandtalab2016non} includes physiological non-EEG signals that are labeled with four neurological stress status. 
The DCNN in \cite{jafari2018sensornet} utilizes a cascade of 5 Conv layers of size 16, 16, 8 and 8 filter banks and 2 FC layers of size 64 and number-of-classes respectively. Input is cut into frames of 64 samples. All filters have a shape of 1$\times$5. We conducted a set of experiments on binarizing this baseline configuration, abiding by the same number of layers, but increasing the number of layer parameters in each experiment. When binarizing the original 16-bit model, the BNN model is compressed by 16 times, but the accuracy drops from the reported 93.8\% to 81.8\%. In order to compensate the accuracy loss to the baseline level, we increased the number of parameters by approximately 9.5$\times$, and still, the final binarized model is 1.7$\times$  smaller than the 16-bit baseline. Fig. \ref{fig:Augmentation} shows the incremental augmentation and the impact of binarizing over the baseline DCNN, and the compensation in accuracy with the increase of binarized model size. In total, the binarized DCNN for Stress Detection has 98K parameters and requires 13KB of memory to store the model. For every classification of this dataset, 7.32 Million operations are required. Fig. \ref{fig:case_models}-A shows the BNN configuration and table \ref{tab:cases} summarizes the details for each case study after augmentation.

\begin{table}[h]
\caption{Detail Summary of the two case studies and the characteristics of their BNNs used in this paper}
\scriptsize
\setlength{\tabcolsep}{6.5pt}

\begin{center}
\resizebox{\columnwidth}{!}{%
\label{my-label}
\begin{tabular}{ccccccc}
\hline
\textbf{Dataset} & \textbf{\begin{tabular}[c]{@{}c@{}}Input\\ Shape\end{tabular}} & \textbf{\begin{tabular}[c]{@{}c@{}}\# of\\ Labels\end{tabular}} & \textbf{\begin{tabular}[c]{@{}c@{}}\# of\\ Params\end{tabular}} & \textbf{\begin{tabular}[c]{@{}c@{}}Model\\ Size\end{tabular}} & \textbf{\begin{tabular}[c]{@{}c@{}}Total \\ Ops\end{tabular}} & \textbf{\begin{tabular}[c]{@{}c@{}}Avg\\ Accuracy\end{tabular}} \\ \hline
Stress & 1$\times$64$\times$7 & 4 & 98K & 13KB & 7.32M & 94.1\% \\ \hline
PAMAP2 & 40$\times$1$\times$1 & 16 & 145K & 18KB & 0.29M & 97.8\% \\ \hline
\end{tabular}%
}
\end{center}
\label{tab:cases}
\end{table}


\subsection{Case study 2: Physical Activity Monitoring}
Physical Activity Monitoring dataset (PAMAP2) \cite{reiss2012introducing} contains data from 9 subjects recorded from sensors such as IMUs, heart rate monitor that totally has 40 valid channels labeled with 12 physical activities. 
We use a 3-layer binarized Multi-Layer Perceptron (MLP) as proposed in \cite{GLSVLSI-ali} for classification of the PAMAP2 dataset with an average accuracy of 97.8\%. In total, the BNN MLP has 145K parameters, storable in 18KB, and requires 290K operations. Fig. \ref{fig:case_models}-B shows the configuration and Table \ref{tab:cases} summarizes the details.

\section{\bf{Scalable Hardware Platform}}
In this section, we introduce a minimal hardware architecture that is scalable with the number of PEs and memory width that can be adjusted to match the time and energy requirements given an application. For the parallelization scheme, we use output channel tiling as in \cite{jetc-page} in which every PE contributes to generating an independent channel of the feature-map. This scheme is shown to result in the best computation to memory communication ratio \cite{zhang2015optimizing}. It also facilitates implementing the Pool-Skipping method with a simple control logic.

\begin{figure}[hbt]
\begin{center}
\includegraphics[width=3in]{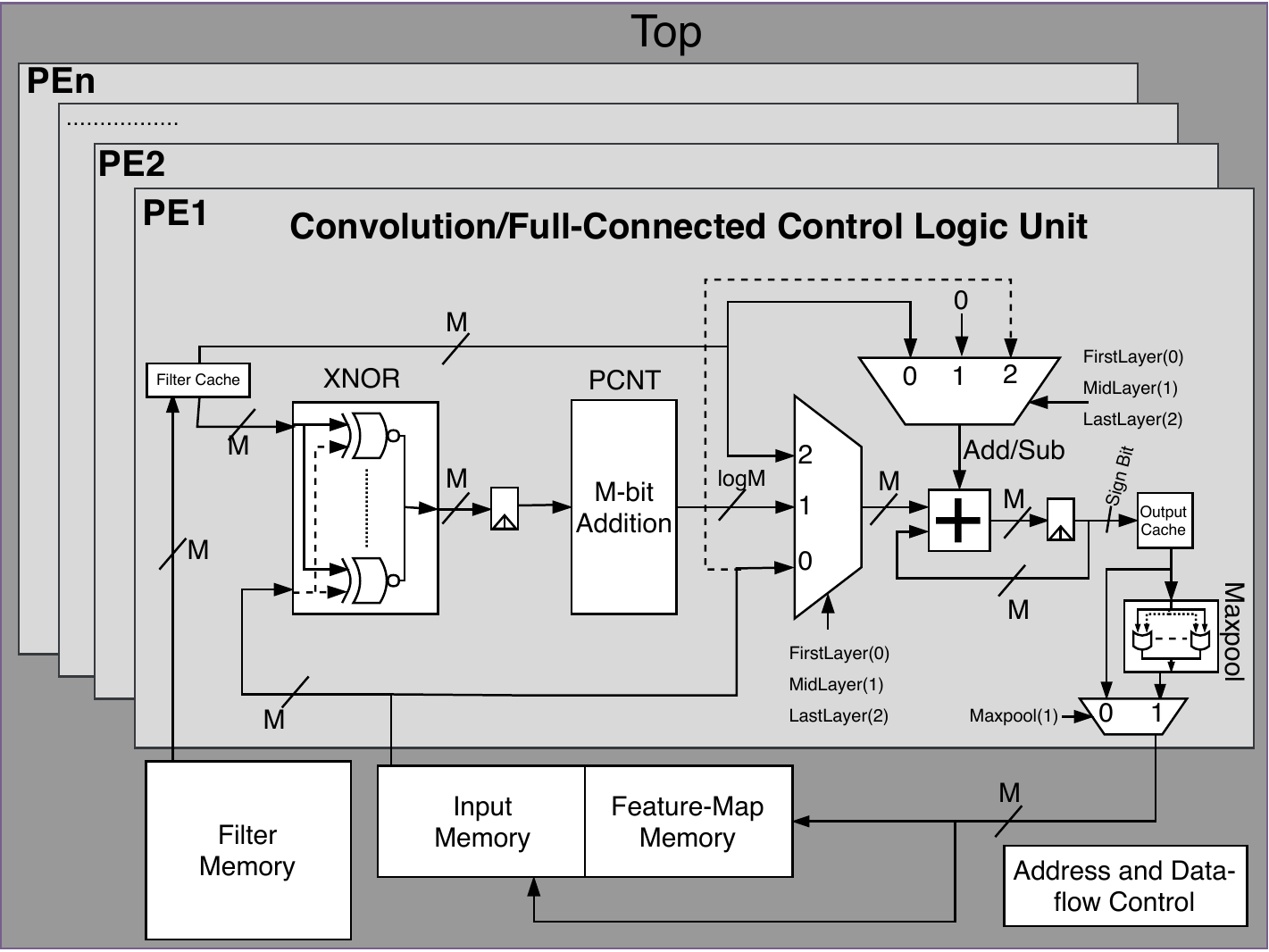}
\end{center}
\caption{Block diagram of hardware architecture which comprises Filter, Input, and Feature-Map memories that are addressed by a global controller, and are shared between PEs. Each PE includes a filter and output cache, a pipeline of bit-wise $xnor$, $pcnt$ and accumulator, and by means of MUX and a local controller performs first/mid/last BNN layer with respect to Conv/FC operations. A bit-wise $or$ gate is also embedded to perform MP. 
}
\label{fig:Sensornet_hardware}
\end{figure}

The high-level abstraction of the hardware is depicted in Fig.\ref{fig:Sensornet_hardware}. The main components of the hardware include Filter Memory to store the model weights, an Input-Memory and a Feature-Map Memory that are used for the intermediate data and alternately switch their tasks, i.e. buffering the input data/feature-map to be exported to PEs in one memory and importing the output feature-map from the PEs in another memory. All the three mentioned memories are shared between the PEs and addressed by a global controller. The global controller is the only unit of the hardware that uses a multiplier for address calculations. 

Each PE comprises a Filter cache to temporarily store an individual filter for the process, and an Output cache to concatenate the single output bits and prepare them to be sent to the Feature-Map Memory along with other PEs. Once this cache is full, all the PEs concatenate their first bit of output-cache and send the packed packet out to the feature-map memory. Then, next bits are sent out accordingly. Resource sharing is carefully taken into consideration such that the major logic of every PE is only a pipeline of bit-wise $xnor$, $pcnt$, and an accumulator to satisfy the equation (\ref{eq:BNNoperation1}). The accumulator ADDs/SUBs over either the input data w.r.t binarized packed filter for the first BNN layer, or on the output of $pcnt$ for mid-layers, or over the filter data w.r.t binarized packed data for the last BNN layer. Two shifters, not depicted in Fig. \ref{fig:Sensornet_hardware} to avoid obfuscation, shift the packed filter/data for the first/last layers. A local controller, by means of Multiplexers to allocate the existing resources, handles the data-flow such that FC and Conv layers with different strides are executed with respect to the BNN layer.

A bit-wise $or$ gate is implemented after the Output cache to perform MP and is by-passed if there is no MP. Pool-Skipping is implemented in such a way that, given MP, whenever there is a +1 in the pool-under-explore, +1 is written quickly to its relevant location in the Output cache, the pipeline is flushed, and the operation is skipped to a stride as wide as the pool size. The would-be adjacent bits in the Output cache are left as `don't care' as their old values will be masked by the effect of the written +1.



\section{\bf{Evaluation}}

\subsection{Parallelization}

As mentioned in Section III, BNN inference performance in our M-bit datawidth machine is proportional to $M$, omitting the first/last layers that may use full-precision values. Meanwhile, with the split and tiled threads dispatched to $N$ concurrent PEs, the computation can be expedited by $N$ times if the memory communication is disregarded. Clock frequency, $f$, is another factor that decides the computation rate. Therefore in our hardware, the performance is proportional to these three metrics:
\begin{equation} 
Performance \propto f \times M \times N
\label{eq:performance}
\end{equation}
The performance of the first and the last layer, similar to traditional accelerators, is proportional to $f$$\times$$N$ only.

\begin{figure}
\begin{center}
\includegraphics[width=2.6in]{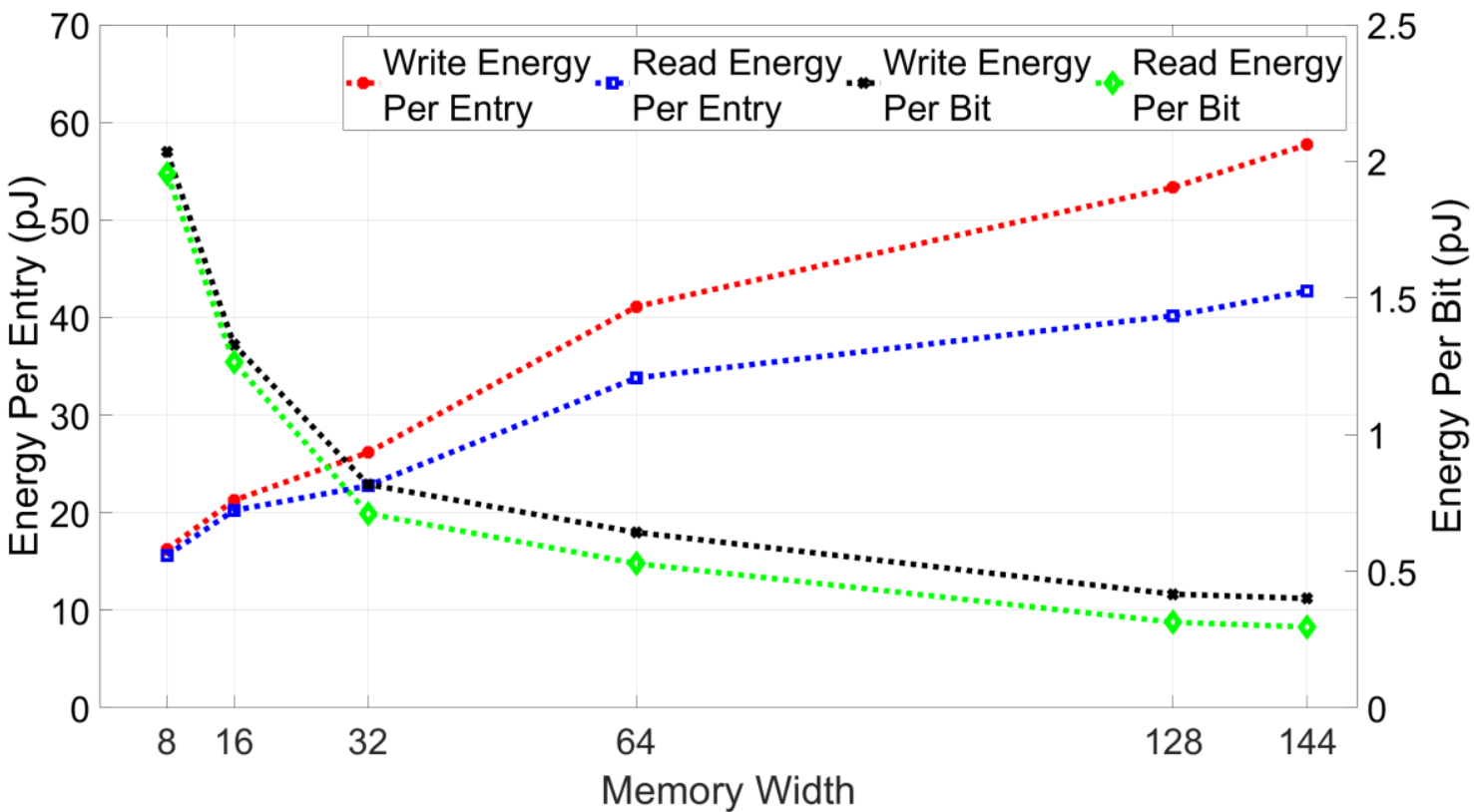}
\end{center}
\caption{Energy curves inferred from the technical data sheets of ARM memories showing the trend for read/write per entry and per bit on increasing ARM Memory Width
\label{fig:MemoryEnergy}}
\end{figure}

\begin{figure}
\begin{center}
\includegraphics[width=2.9in]{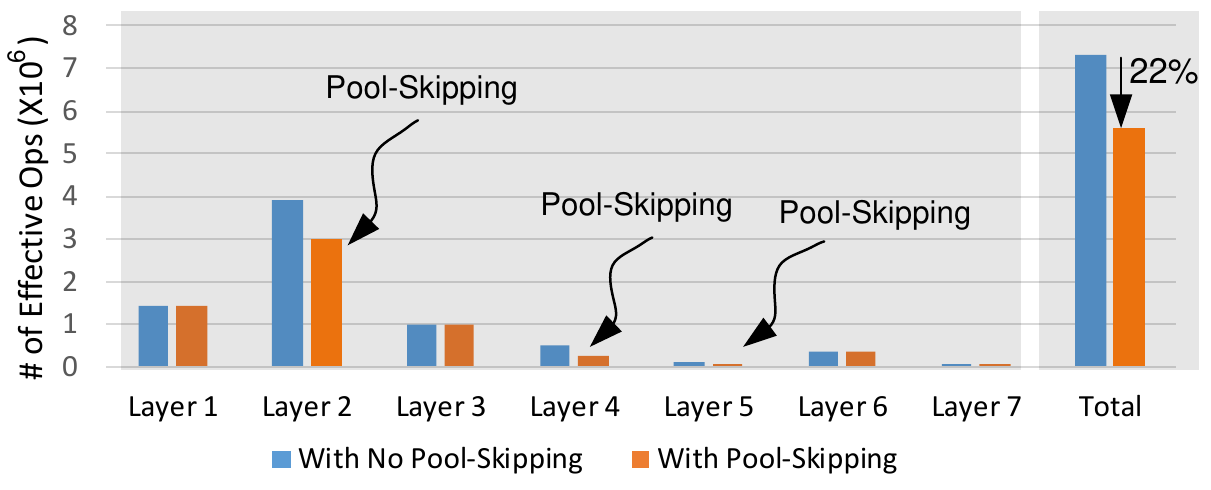}
\end{center}
\caption{Impact of Pool-Skipping on the number of effective operations for the Stress Detection case study 
\label{fig:Pool-Skipping}}
\end{figure}
\subsection{Memory Energy Evaluation}
We will seek a trade-off between $M$ and $N$ in the next section for minimizing the energy consumption, but beforehand, we analyze a set of ARM memories generated with Artisan Memory Compiler in 65nm and at frequency of 128MHz with various datawidth. For each generated memory we registered the energy per entry read/write, and calculated the energy per bit read/write. Fig. \ref{fig:MemoryEnergy} plots the results of this analysis and indicates that it is more energy efficient to read 1 bit of information, as a BNN weight, if wider memories are used.

\subsection{Pool-Skipping}
\label{sub-sec:PoolSkipping}
As explained in Section III, Pool-Skipping can skip a large portion of operations that are preceding a MP. For the Stress Detection case study, 3 MPs with pool size of 1$\times$2 follow 3 Conv layers that contribute to 76.5 \% of the total operations. Fig. \ref{fig:Pool-Skipping} shows the impact of Pool-Skipping on the number of operations per layer and in total for the BNN of Stress Detection. Applying Pool-Skipping, the average number of effective operations is reduced by 22\%.


\section{\bf{Implementation Results and Analysis}}
\subsection{FPGA Implementation}
\begin{figure}%
    \centering
    \includegraphics[width = 4.4 cm]{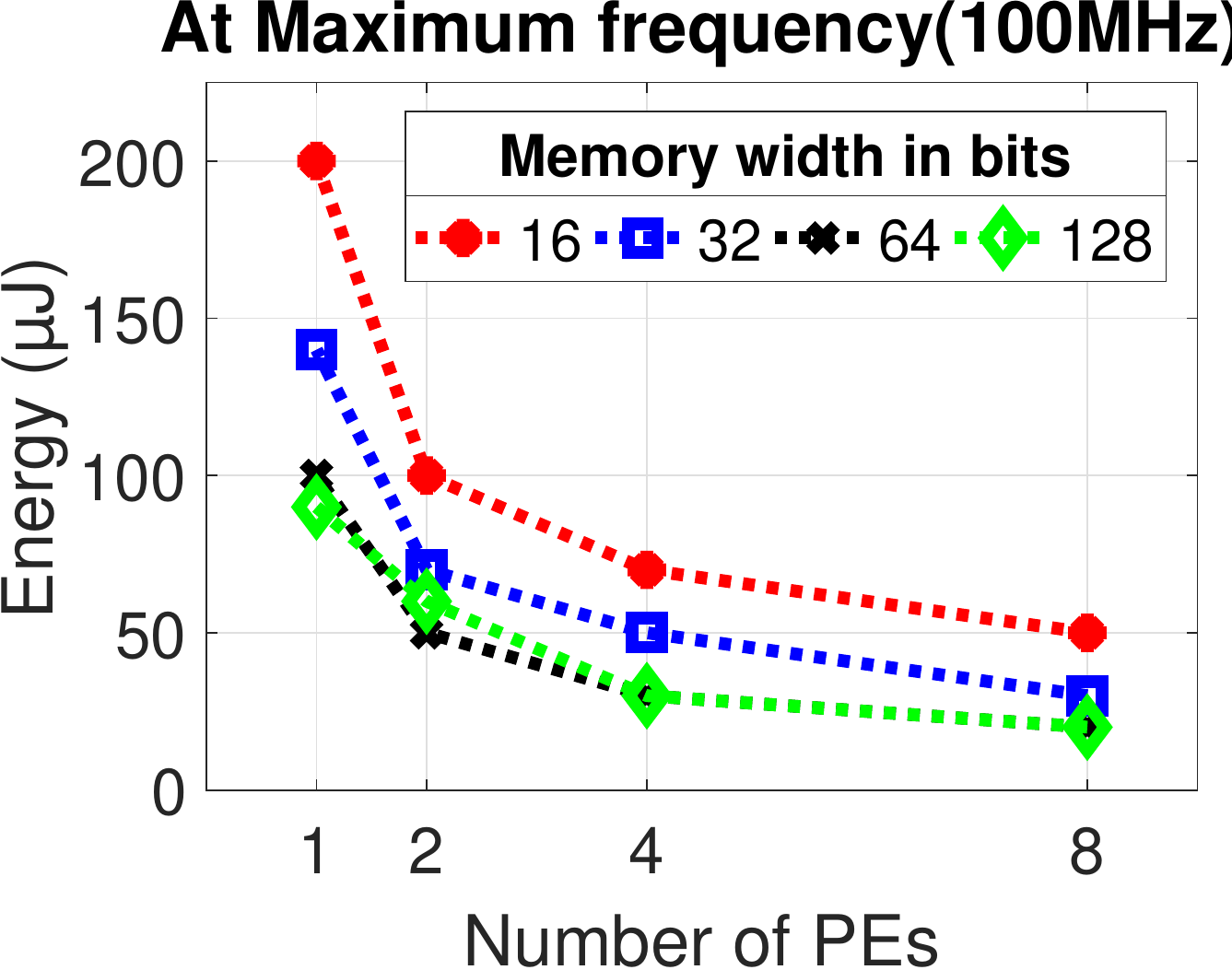}%
    \qquad
    \hspace{-5ex}
    \includegraphics[width = 4.4 cm]{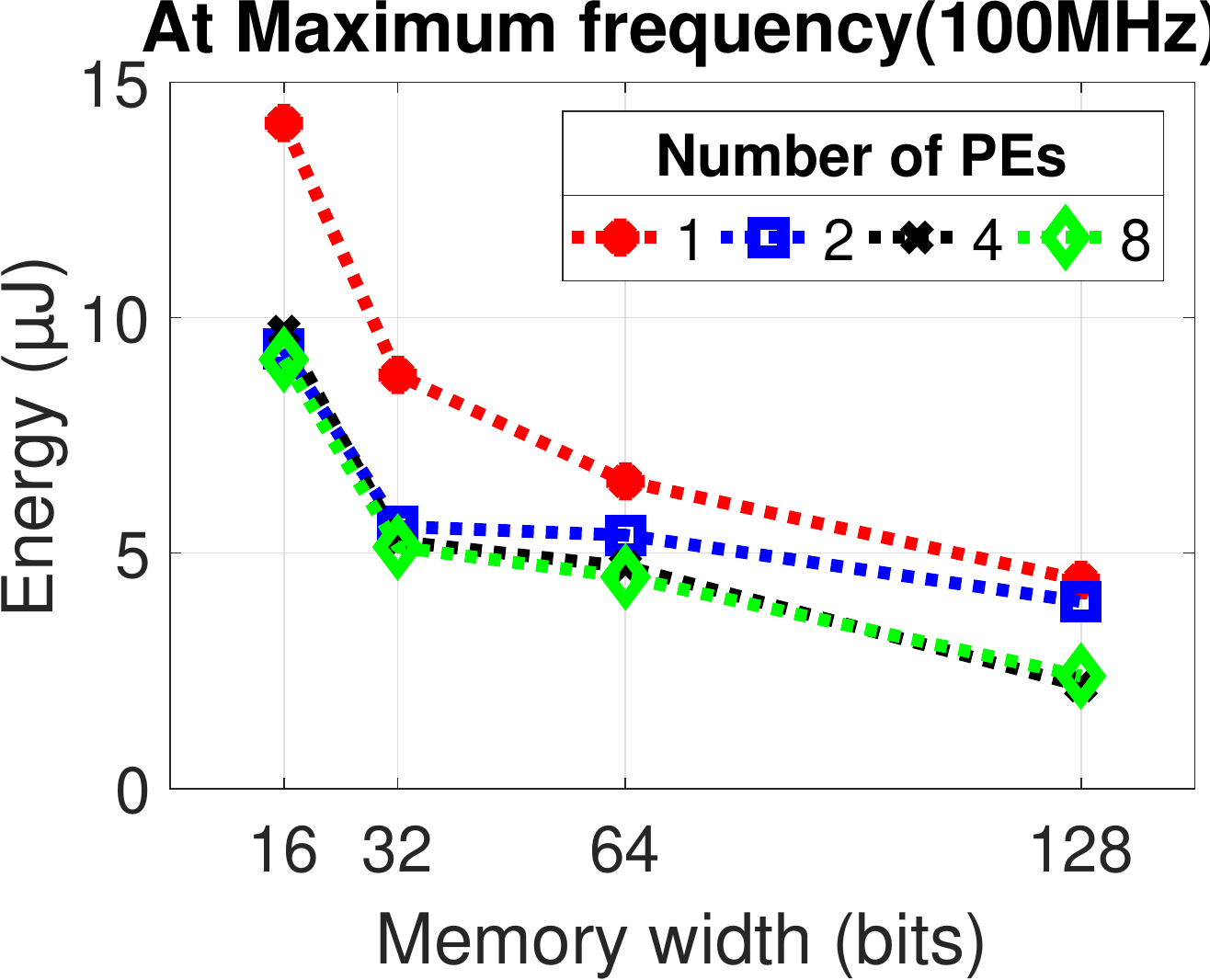}%
    \caption{(left)  impact on the classification energy with the increase of the number of PEs for four different Memory widths for the Stress Detection (right) impact on classification energy with increasing the number of Memory width for 4 different PEs for the Physical Activity Monitoring on the FPGA }%
    \label{fig:fpga}%
\end{figure}

In order to emulate the case studies on the proposed hardware, FPGA is used for the evaluation setup. We choose the low-power Artix-7 family and select the smallest package whose Block RAMs (BRAMs) can suffice the model of the case studies. Each case study was synthesized, placed and routed and implemented using Xilinx Vivado. with different numbers of PEs and different width for BRAMs. Fig. \ref{fig:fpga}-(left) depicts the classification energy for the Stress Detection vs number of PEs w.r.t varying memory widths, and Fig. \ref{fig:fpga}-(right) depicts classification energy vs Memory width with varying number of PEs for Physical Activity. Table II provides the implementation results for the two cases with memory width of 128 bits, at 100MHz and with more implementation details. 
In Table II, for each case study the optimal (Opt.) configuration in terms of energy is given for a specific number of PEs. For the Stress Detection, the optimal config. (8PEs) requires 4$\times$ less energy, and for the Physical Activity, the optimal config.(4PEs) consumes 2$\times$ less energy as compared to their serial configs (1PE).





\begin{table}
\caption{Implementation results of the proposed BNN processor on Xilinx FPGA (Artix-7). The Results obtained at the clock frequency of 100 MHz and with Memory width of 128 bits}
\scriptsize
\setlength{\tabcolsep}{1.5pt}
\begin{center}
\begin{tabular}{|c|c|c|c|c|c|c|}
\hline
\multirow{2}{*}{\begin{tabular}[c]{@{}c@{}}Case Studies\\ \\ Implementation\\ Characteristics\end{tabular}} & \multicolumn{3}{c|}{\begin{tabular}[c]{@{}c@{}}Stress \\ Detection\end{tabular}} & \multicolumn{3}{c|}{\begin{tabular}[c]{@{}c@{}}Physical Activity \\ Monitoring\end{tabular}} \\ \cline{2-7} 
 & Serial & \begin{tabular}[c]{@{}c@{}}Semi \\ Parallel\end{tabular} & \begin{tabular}[c]{@{}c@{}}Fully \\ Parallel\\ (Opt.)\end{tabular} & Serial & \begin{tabular}[c]{@{}c@{}}Semi \\ Parallel\\ (Opt.)\end{tabular} & \begin{tabular}[c]{@{}c@{}}Fully \\ Parallel\end{tabular} \\ \hline
\# of PEs & 1 & 4 & 8 & 1 & 4 & 8 \\ \hline
BRAM & 6 & 6 & 6 & 3 & 3 & 3 \\ \hline
DSP & 1 & 1 & 1 & 1 & 1 & 1 \\ \hline
\# of slices & 341 & 1076 & 2047 & 385 & 1329 & 2601 \\ \hline
Latency (ms) & 1.14 & 0.31 & 0.19 & 0.06 & 0.03 & 0.03 \\ \hline
Throughput(k label/s) & 0.87 & 3.23 & 5.22 & 17.17 & 36.15 & 38.12 \\ \hline
Total Power (mW) & 78 & 98 & 115 & 76 & 79 & 91 \\ \hline
Energy (µuJ) & 89.3 & 30.3 & 22.0 & 4.4 & 2.2 & 2.4 \\ \hline
\end{tabular}
\end{center}
\label{tab:FPGA_results}
\end{table}

\subsection{ASIC Implementation}
\begin{figure}%
    \centering

    \includegraphics[width = 4.4 cm]{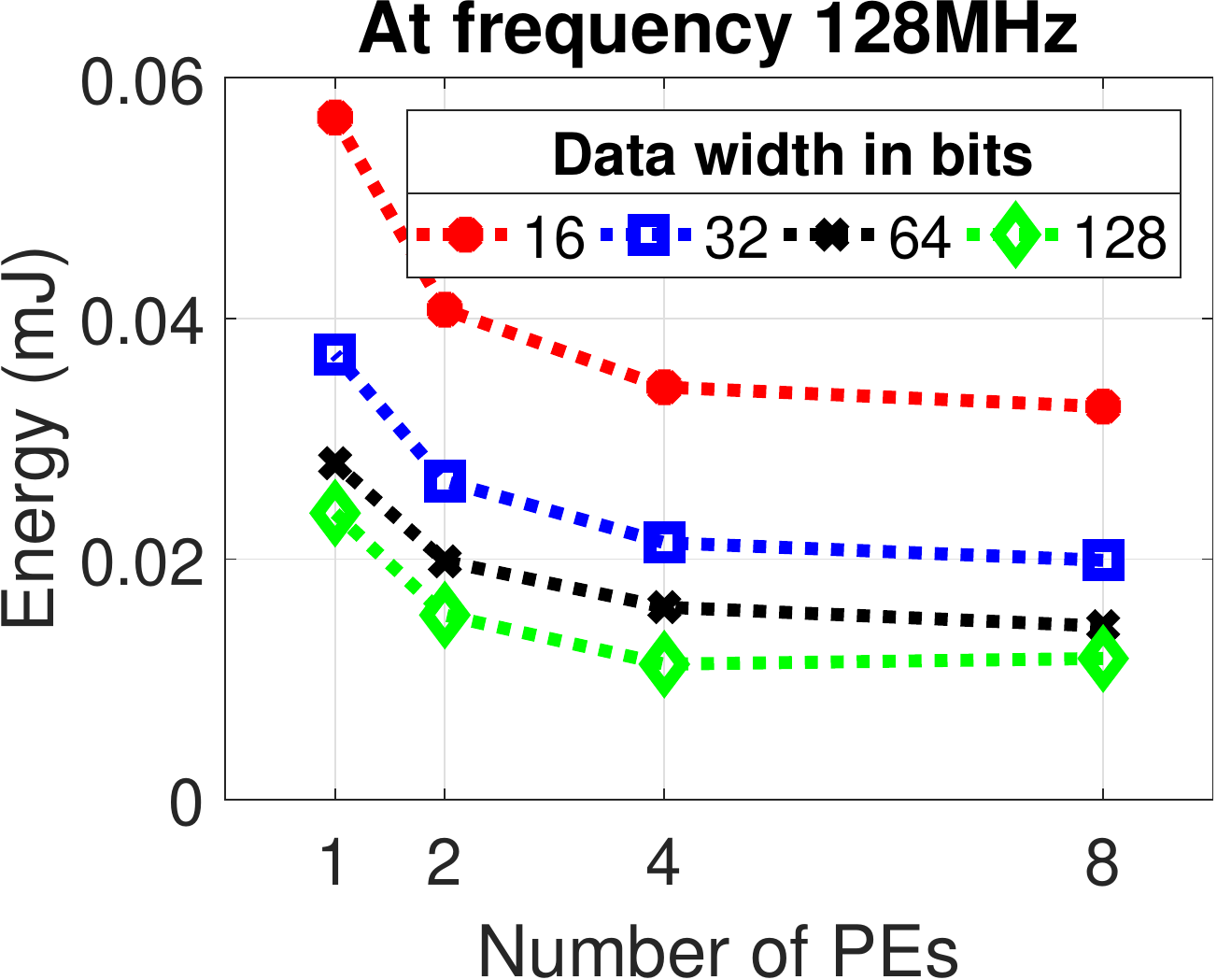}%
    \qquad
    \hspace{-5ex}
    \includegraphics[width = 4.4 cm]{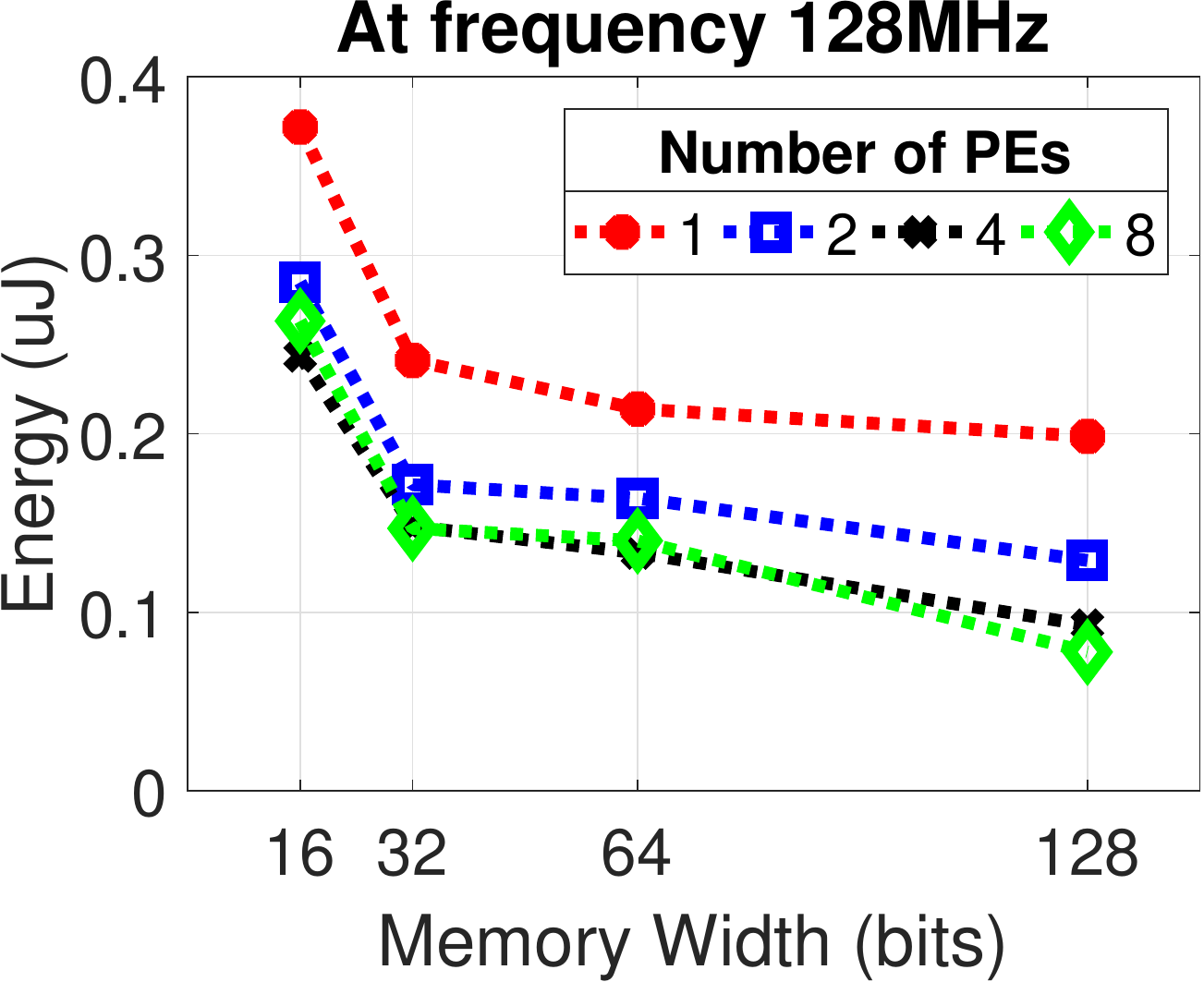}%
    \caption{(left)  impact on classification energy with increase of the number of PEs for four different Memory widths for Stress Detection (right) impact on classification energy with increasing the Memory widths for 4 different PEs for Physical Activity on a 65nm ASIC}%
    \label{fig:asic}%
\end{figure}

\begin{figure}[t]
\begin{center}
\includegraphics[width=3.3in]{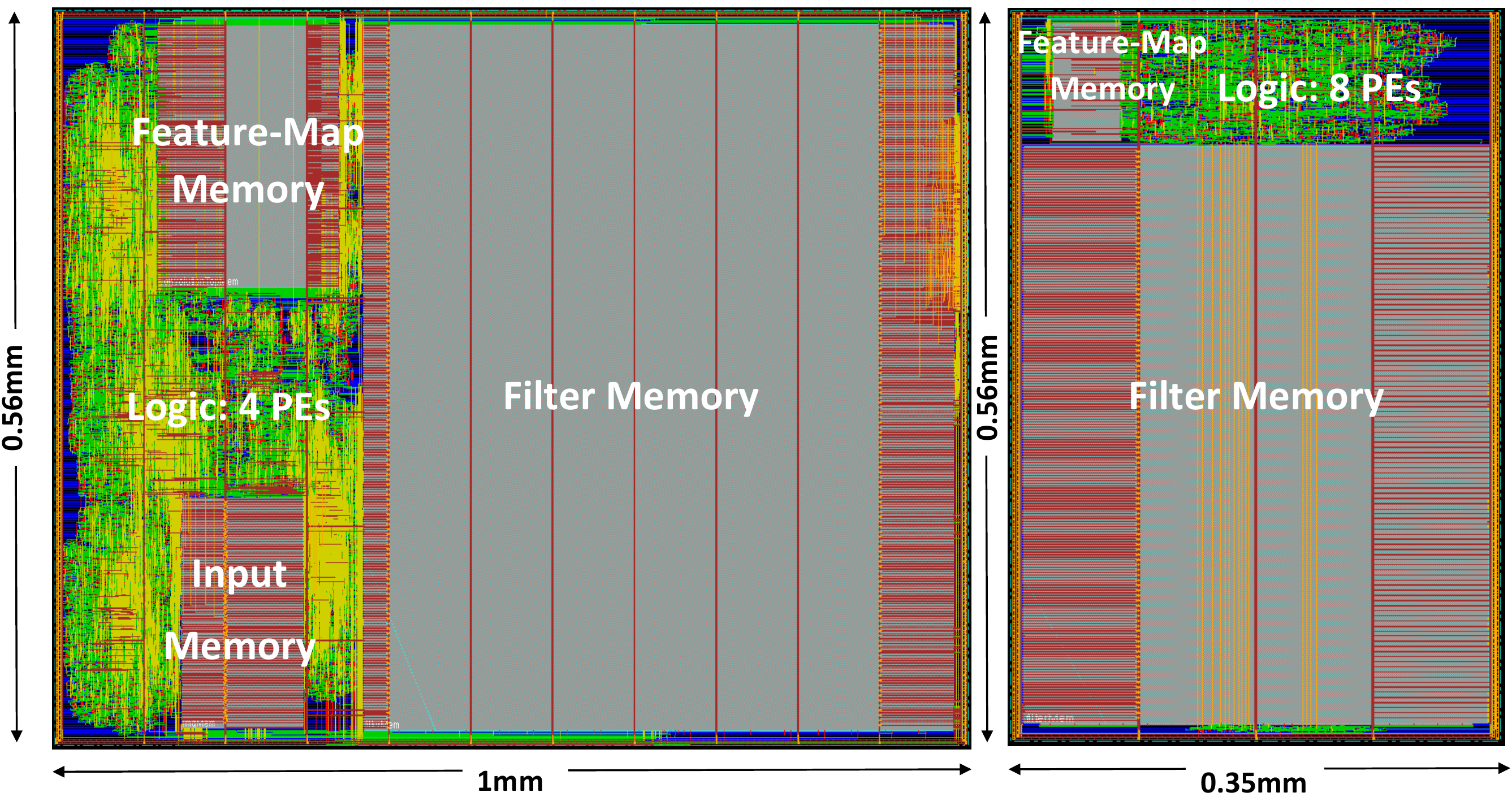}
\end{center}
\caption{Post-layout views of BNN for Stress Detection (left) and for Physical Activity (right), in 65nm TSMC CMOS technology.}
\label{fig:layout}
\end{figure}

Various configurations per case study were synthesized using RC Compiler and post place-and-routed using the Encounter SOC from Cadence using 65nm standard cell library CMOS technology. 
For the two case studies we obtained slightly different optimal configs per case study on ASIC. Fig. \ref{fig:asic}-(left) and Fig. \ref{fig:asic}-(right) respectively depict the classification energy for the first and second case studies w.r.t. the varying PEs and varying memory widths, and Table III provides the implementation results with memory width 128 bits, at 128MHz in more details. Both the Figure and the table confirm energy per bit analysis inferred from Fig. \ref{fig:MemoryEnergy}. From table III it is concluded that for the Stress Detection and Physical Activity, the optimal configs are respectively 2$\times$ and 2.5$\times$ more energy efficient than their serial config counterpart. The ASIC post-layout views for the optimal configs of both cases are shown in Fig. \ref{fig:layout}. The reason for the Physical Activity case to have smaller PE logic is that in its HDL the MP and FC logics are automatically removed and less number of SRAM cells have been dedicated as the cache memories.




\begin{table}
\caption{post place and route results of the proposed BNN processor on ASIC CMOS 65nm 1.1v. The Results obtained at clock frequency of 128 MHz and with a memory width of 128 bits}
\scriptsize
\setlength{\tabcolsep}{1.5pt}
\begin{center}
\begin{tabular}{|c|c|c|c|c|c|c|}
\hline
\multirow{2}{*}{\begin{tabular}[c]{@{}c@{}}Case Studies\\ \\ Implementation\\ Characteristics\end{tabular}} & \multicolumn{3}{c|}{\begin{tabular}[c]{@{}c@{}}Stress\\ Detection\end{tabular}} & \multicolumn{3}{c|}{\begin{tabular}[c]{@{}c@{}}Physical Activity\\ Monitoring\end{tabular}} \\ \cline{2-7} 
 & Serial & \begin{tabular}[c]{@{}c@{}}Semi \\ Parallel\\ (Opt.)\end{tabular} & \begin{tabular}[c]{@{}c@{}}Fully \\ Parallel\end{tabular} & Serial & \begin{tabular}[c]{@{}c@{}}Semi \\ Parallel\end{tabular} & \begin{tabular}[c]{@{}c@{}}Fully \\ Parallel\\ (Opt.)\end{tabular} \\ \hline
\# of PEs & 1 & 4 & 8 & 1 & 4 & 8 \\ \hline
Area Util. (\%) & 94 & 91 & 90 & 94 & 93 & 92 \\ \hline
Clock Freq. (MHz) & \multicolumn{6}{c|}{128} \\ \hline
Max Freq. (MHz) & \multicolumn{6}{c|}{1000} \\ \hline
Core Area (mm2) & 0.49 & 0.56 & 0.66 & 0.18 & 0.19 & 0.20 \\ \hline
Latency (ms) & 2.13 & 0.55 & 0.29 & 0.04 & 0.018 & 0.014 \\ \hline
Throughput(k label/s) & 0.47 & 1.80 & 3.42 & 22.22 & 52.81 & 68.52 \\ \hline
Leakage power (mW) & 5.77 & 6.42 & 7.27 & 3.61 & 3.66 & 3.71 \\ \hline
Dynamic power (mW) & 5.37 & 13.83 & 32.85 & 0.80 & 1.23 & 1.62 \\ \hline
Total power (mW) & 11.14 & 20.25 & 40.13 & 4.41 & 4.90 & 5.32 \\ \hline
Energy (uJ) & 23.81 & 11.25 & 11.73 & 0.20 & 0.09 & 0.077 \\ \hline


\end{tabular}
\end{center}
\label{tab:ASIC_results}
\end{table}

\section{\bf{Comparison}}
We compare our proposed hardware on the two case studies with two relevant works, \cite{jafari2018sensornet} and \cite{GLSVLSI-ali}, from which we obtained our BNN configurations. The comparison is summarized in Table IV. In summary, despite 9.5$\times$ augmentation in the number of parameters, the optimal config. for the Stress Detection is 4.5$\times$ and 6.8$\times$ more energy efficient as compared to \cite{jafari2018sensornet} implemented respectively on Artix-7 FPGA and on ASIC 65nm. For the Physical Activity, the minimal hardware with no MP and Conv logics, is 250$\times$ more energy efficient as compared to a programmable, yet low-power manycore platform in \cite{GLSVLSI-ali}. \\

\begin{table}[h]
\textbf{\caption{Comparison of the proposed hardware with optimized configs with related works}}
\scriptsize
\setlength{\tabcolsep}{3.2pt}
\begin{center}
\begin{tabular}{|c|c|c|c|c|c|c|}
\hline
Metrics & \multicolumn{2}{c|}{\cite{jafari2018sensornet}} & \multicolumn{2}{c|}{\textbf{This Work}} & \cite{GLSVLSI-ali} & \textbf{This Work} \\ \hline
Application & \multicolumn{4}{c|}{Stress Detection} & \multicolumn{2}{c|}{Physical Activity} \\ \hline
Technique & \multicolumn{2}{c|}{16-bit DCNN} & \multicolumn{2}{c|}{BNN} & BNN & BNN \\ \hline
Accuracy(\%) & \multicolumn{2}{c|}{94} & \multicolumn{2}{c|}{94.1} & 97.8 & 97.8 \\ \hline
Platform & \begin{tabular}[c]{@{}c@{}}FPGA\\ Artix-7\end{tabular} & \begin{tabular}[c]{@{}c@{}}ASIC \\ 65nm\end{tabular} & \begin{tabular}[c]{@{}c@{}}FPGA\\ Artix-7\end{tabular} & \begin{tabular}[c]{@{}c@{}}ASIC\\ 65nm\end{tabular} & \begin{tabular}[c]{@{}c@{}}ASIC \\ 65nm\end{tabular} & \begin{tabular}[c]{@{}c@{}}ASIC \\ 65nm\end{tabular} \\ \hline
Freq. (MHz) & 100 & 100 & 100 & 128 & 145 & 128 \\ \hline
Latency (us) & 1000 & 800 & 190 & 550 & 77 & 14 \\ \hline
Power (mW) & 154 & 60 & 115 & 22 & 270 & 5.3 \\ \hline
Energy (uJ) & 150 & 50 & 22 & 11 & 20 & 0.08 \\ \hline
Energy Imp. & - & - & \textbf{6.8$\times$} & \textbf{4.5$\times$} & - & \textbf{250$\times$} \\ \hline
\end{tabular}
\end{center}
 \label{tab:HW_comparison_related}
\end{table}

\section{\bf{Conclusion}}
We proposed a scalable hardware for inference of BNNs that have been trained for physiological datasets. The proposed hardware has flexibility in adjusting the memory width and number of PEs, both adjustable for the most energy efficient configuration while meeting time requirements given an application. Two case studies including Physical Activity Monitoring and Stress Detection are trained and deployed on the hardware, and for each, the optimal number of PEs and memory-width is sought. Our implementation results with the two case studies on Artix-7 FPGA and on 65nm ASIC CMOS standard cell library indicate that the configuration parameters of the hardware, if adjusted well, can optimize the energy consumption up to 4$\times$ and 2.5$\times$ respectively on FPGA and on ASIC. An operation skipping method, named Pool-Skipping, is also proposed for BNNs that skips effectless operations that precede a Max-Pool layer, and can skip up to 22$\%$ of operations in the Stress Detection case study. When compared to the related works that use the same case studies on the same target platforms, FPGA and ASIC, and with the same classification accuracy, our hardware is respectively 4.5$\times$ and 250$\times$ more energy efficient for the Stress Detection on FPGA and Physical Activity Monitoring on ASIC.







\bibliographystyle{IEEEtran}
\input{ISQEDsamp.bbl}

\end{document}